\documentclass[a4paper,12pt]{article}
\usepackage{eurosym}
\usepackage[dvips]{graphicx}
\usepackage{amsfonts}
\usepackage{mathrsfs}
\usepackage{graphicx} %We can use any other package if it is necessary
\usepackage{epsfig}
\usepackage{amsmath, amsthm}
\usepackage{amssymb}

\usepackage{amsmath}
\usepackage{amssymb}
\usepackage{mathrsfs}
\usepackage{amsthm}
\usepackage{graphicx}
\usepackage{color}
\usepackage{float}
\usepackage{subfigure}
\usepackage{framed}
\restylefloat{table}
\newcommand{\ba}{\begin{array}}
\newcommand{\ea}{\end{array}}

\newcommand{\beq}{\begin{equation}}
\newcommand{\eeq}{\end{equation}}
\newcommand{\ben}{\begin{enumerate}}
\newcommand{\een}{\end{enumerate}}
\newcommand{\bit}{\begin{itemize}}
\newcommand{\eit}{\end{itemize}}

\newtheorem{theorem}{Theorem}

\begin{document}
\title{General rogue wave solutions to the discrete nonlinear Schr\"odinger equation}
\author{Yasuhiro Ohta \\
\textsl{Department of Mathematics, Kobe University}\\
\textsl{Rokko, Kobe 657-8501, Japan}\\
\and
Bao-Feng Feng \\
%EndAName
\textsl{School of Mathematical and Statistical Sciences} \\
\textsl{The University of Texas Rio Grande Valley}}
\maketitle

\begin{abstract}
In the present paper, we attempt to construct both the general rogue wave solutions to the fully discrete  nonlinear Schr\"odinger (fd-NLS) equation via the KP-Toda reduction method. 
First, we deduce the general breather solution of the fd-NLS equation starting from a pair of bilinear equations. We then derive the general rogue wave solution by taking a limit to the breather solution.
\end{abstract}
%\newpage
\section{Introduction}
Rogue waves (RWs) or freak waves are spontaneously excited local nonlinear waves with large amplitudes  which appear from nowhere and disappear with no trace \cite{akhmediev2009waves}.
The simplest form of such waves were firstly discovered Peregrine in the nonlinear Schr\"odinger (NLS) equation \cite{peregrine1983water}, and their higher order forms were found 
20 years later in \cite{akhmediev2009rogue,kedziora2012second,dubard2010multi,dubard2011multi,gaillard2011families,guo2012nonlinear,ohta2012general}. 
Such extreme wave have been observed in various different contexts such as
oceanography \cite{kharif2009rogue}, hydrodynamic \cite{onorato2013rogue,chabchoub2011rogue}, Bose-Einstein condensate \cite{bludov2009matter}, plasma \cite{bailung2011observation} and nonlinear optic \cite{onorato2013rogue,solli2007optical,kibler2010peregrine}.

Motivated by these physical applications, rogue wave solutions have been found in many other nonlinear wave equations such as the derivative Schr\"odinger (NLS) equation \cite{HedNLS2011,GuodNLS2013,ChowdNLS2014,HedNLS2017,JiankeJunchaoJNS}, the Manakov system \cite{BaroniocNLS2012,BaroniocNLS2014,YYManakov}, Davey-Stewartson I and II equation \cite{OhtaYangDSI,OhtaYangDSII}, the three-wave equation \cite{JiankeBoIMA}, the Boussinesq equation \cite{JiankeBoussinesq}, the Yajima-Oikawa equation \cite{chen2015rational,chen2017YORW},.

On the other hand,  the study of rogue waves in discrete integrable systems is much less. As far as we are aware, only the rogue waves in semi-discrete NLS equation, or the so-called Ablowitz-Ladik equation have been reported in the literature \cite{ankiewicz2010discrete,ohta2014general}. In the present paper, we attempt to construct the rogue wave solutions to the fully discrete nonlinear Schr\"odinger  (fd-NLS) equation, which was
originally discovered by Ablowitz and Ladik \cite{AL1,AL2} and rediscovered by Hirota and Ohta through Hirota's bilinear approach \cite{Dis-NLS}.  It was also discussed in \cite{Dis-NLS-Tsujimoto1,Dis-NLS-Tsujimoto2}. In a recent paper by Hirose et al., the integrable discretization of the local induction equation which is gauge equivalent to the NLS equation was presented in \cite{ILE2019}.
 
 The remainder of the present paper is organized as follows. In section 2, we will construct general breather solution to the fd-NLS equation. We derive general rogue wave solutions to the fd-NLS equation by taking the limit to the general breather solution in section 3. The paper is concluded in section 4. 
%Fundamental rogue wave solution can be obtained from one-breather solutions by taking $+$ to $-$ and taking the limit of $p_i \to 0$.
%\begin{eqnarray*}
%\tau _{n}^{s}(l)
%&=& \left[ (r+1)l+ (r-1)(n-n_0) + (r-1+r\bar{c}-c^{-1})(s-s_0) \right] \\
%&& \left[-(r+1) l+ (r-1)(n-n_0) + (r-1+rc-\bar{c}^{-1})(s-s_0) \right] + r
%\end{eqnarray*}

\section{Breather solution for fully discrete NLS equation}
Based on the bilinear formulation, we are able to derive the breather solution of the fully discrete NLS equation, which is given by the following theorem.
\begin{theorem}
The fully discrete NLS equation,
\begin{equation}
\left\{
\begin{array}{l}
\mathrm{i}(q_k^{t+1}-q_k^t)
=(q_{k+1}^t+q_{k-1}^{t+1})(1+\epsilon |q_k^t|^2)\Gamma_k^t, \\[5pt]
\displaystyle
\Gamma_{k+1}^t=\frac{1+\epsilon |q_k^t|^2}{1+\epsilon |q_k^{t+1}|^2}\Gamma_k^t,
\end{array}
\right.
\label{dNLS}
\end{equation}
where $\epsilon=\pm 1$, admits the $N$-breather solution,
\begin{equation}
q_k^t=\frac{g_k^t}{f_k^t}\frac{1-r}{2\sqrt{|r|}}e^{\mathrm{i}k\theta}
\left(\frac{1-c}{1-\bar c}\frac{1+r\bar c}{1+rc}\right)^t,\quad
\Gamma_k^t=\frac{4rR}{(1+r)^{2}}
\frac{f_{k-1}^{t+1}f_k^t}{f_k^{t+1}f_{k-1}^{t}},
\label{vartr}
\end{equation}
where $\bar{\ }$ means complex conjugate,
$r$ is a real constant whose sign coincides with $\epsilon$,
$\epsilon=\hbox{sign}\,r$, $c$ is a complex constant,
$R$ and $\theta$ are determined by
\begin{equation}
\frac{1}{\mathrm{i}}\frac{1+c}{1-\bar c}\frac{1-r\bar c}{1+rc}
=Re^{\mathrm{i}\theta},
\label{Rth}
\end{equation}
and $f_k^t$ and $g_k^t$ are given by
\begin{equation}
f_k^t=\tau_k^t(0),\quad
g_k^t=\tau_k^t(1),
\label{fg}
\end{equation}
with
\begin{equation}
\tau_k^t(n)=\det_{1\le i,j\le N}\left(A_{ij}^{(n)}(k,t)\right),
\end{equation}
\begin{eqnarray}
&&A_{ij}^{(n)}=\frac{1}{1-rp_i\bar p_j}
\left(a_i\varphi_n(p_i)\overline{a_j\varphi_{-n}(p_j)}
+b_i\varphi_n(-p_i)\overline{b_j\varphi_{-n}(-p_j)}\right)  \nonumber \\
&&\qquad -\frac{1}{1+rp_i\bar p_j}
\left(a_i\varphi_n(p_i)\overline{b_j\varphi_{-n}(-p_j)}
+b_i\varphi_n(-p_i)\overline{a_j\varphi_{-n}(p_j)}\right),
\end{eqnarray}
\begin{equation}
\varphi_n(p)=\left(\frac{1-rp}{1+p}\right)^n\left(\frac{1+p}{1+rp}\right)^k
\left(\frac{1+p}{1+rp}\frac{1+p/\bar c}{1+rcp}\right)^t,
\label{varphi}
\end{equation}
where $p_i$, $a_i$, $b_i$ are complex constants.
\end{theorem}
\noindent
\begin{proof}
%Now we give a brief proof of Theorem 1.
By using the Gramian technique, we can directly prove that
the $\tau$ function of discrete KP/Toda hierarchy,
\begin{equation*}
\tau_n(k,K,l,L)=\det_{1\le i,j\le N}\left(m_{ij}^{(n)}(k,K,l,L)\right),
\end{equation*}
satisfies the discrete bilinear equations,
\begin{equation}
\left\{
\begin{array}{l}
(1-aA)\tau_n(k+1,K+1,l,L)\tau_n(k,K,l,L) \\[5pt]
\quad -\tau_n(k+1,K,l,L)\tau_n(k,K+1,l,L) \\[5pt]
\quad +aA\tau_{n+1}(k+1,K,l,L)\tau_{n-1}(k,K+1,l,L)=0, \\[5pt]
A(a-b)(1-aB)\tau_{n+1}(k+1,K,l+1,L+1)\tau_n(k,K+1,l,L) \\[5pt]
\quad -a(A-B)(1-Ab)\tau_{n+1}(k+1,K,l,L)\tau_n(k,K+1,l+1,L+1) \\[5pt]
\quad +Ab(1-aA)\tau_{n+1}(k,K,l+1,L+1)\tau_n(k+1,K+1,l,L) \\[5pt]
\quad -aB(1-aA)\tau_{n+1}(k+1,K+1,l,L)\tau_n(k,K,l+1,L+1)=0,
\end{array}
\right.
\label{bilin1}
\end{equation}
if the matrix element $m_{ij}^{(n)}$ satisfies
\begin{eqnarray*}
&&m_{ij}^{(n+1)}(k,K,l,L)-m_{ij}^{(n)}(k,K,l,L)
=(-1)^{n+1}\phi_i^{(n)}(k,K,l,L)\psi_j^{(-n-1)}(k,K,l,L), \\
&&m_{ij}^{(n)}(k+1,K,l,L)-m_{ij}^{(n)}(k,K,l,L)
=(-1)^na\phi_i^{(n)}(k+1,K,l,L)\psi_j^{(-n)}(k,K,l,L), \\
&&m_{ij}^{(n)}(k,K+1,l,L)-m_{ij}^{(n)}(k,K,l,L)
=(-1)^nA\phi_i^{(n-1)}(k,K+1,l,L)\psi_j^{(-n-1)}(k,K,l,L), \\
&&m_{ij}^{(n)}(k,K,l+1,L)-m_{ij}^{(n)}(k,K,l,L)
=(-1)^nb\phi_i^{(n)}(k,K,l+1,L)\psi_j^{(-n)}(k,K,l,L), \\
&&m_{ij}^{(n)}(k,K,l,L+1)-m_{ij}^{(n)}(k,K,l,L)
=(-1)^nB\phi_i^{(n-1)}(k,K,l,L+1)\psi_j^{(-n-1)}(k,K,l,L),
\end{eqnarray*}
where $\phi_i^{(n)}$ and $\psi_j^{(n)}$ are arbitrary functions
satisfying the linear dispersion relations,
\begin{eqnarray*}
&&\phi_i^{(n)}(k,K,l,L)-\phi_i^{(n)}(k-1,K,l,L)=a\phi_i^{(n+1)}(k,K,l,L), \\
&&\phi_i^{(n)}(k,K,l,L)-\phi_i^{(n)}(k,K-1,l,L)=A\phi_i^{(n-1)}(k,K,l,L), \\
&&\phi_i^{(n)}(k,K,l,L)-\phi_i^{(n)}(k,K,l-1,L)=b\phi_i^{(n+1)}(k,K,l,L), \\
&&\phi_i^{(n)}(k,K,l,L)-\phi_i^{(n)}(k,K,l,L-1)=B\phi_i^{(n-1)}(k,K,l,L), \\
&&\psi_j^{(n)}(k+1,K,l,L)-\psi_j^{(n)}(k,K,l,L)=a\psi_j^{(n+1)}(k,K,l,L), \\
&&\psi_j^{(n)}(k,K+1,l,L)-\psi_j^{(n)}(k,K,l,L)=A\psi_j^{(n-1)}(k,K,l,L), \\
&&\psi_j^{(n)}(k,K,l+1,L)-\psi_j^{(n)}(k,K,l,L)=b\psi_j^{(n+1)}(k,K,l,L), \\
&&\psi_j^{(n)}(k,K,l,L+1)-\psi_j^{(n)}(k,K,l,L)=B\psi_j^{(n-1)}(k,K,l,L),
\end{eqnarray*}
where $k$, $K$, $l$, $L$ are discrete independent variables
and $a$, $A$, $b$, $B$ are difference intervals.
In order to derive the breather solution, we assume
\begin{eqnarray*}
&&m_{ij}^{(n)}(k,K,l,L) \\
&&\quad
=\sum_{\nu=1}^2\sum_{\mu=1}^2\frac{a_{i\nu}b_{j\mu}}{p_{i\nu}+q_{j\mu}}
\left(-\frac{p_{i\nu}}{q_{j\mu}}\right)^n
\left(\frac{1+aq_{j\mu}}{1-ap_{i\nu}}\right)^k
\left(\frac{1+A/q_{j\mu}}{1-A/p_{i\nu}}\right)^K
\left(\frac{1+bq_{j\mu}}{1-bp_{i\nu}}\right)^l
\left(\frac{1+B/q_{j\mu}}{1-B/p_{i\nu}}\right)^L, \\
&&\phi_i^{(n)}(k,K,l,L)=\sum_{\nu=1}^2a_{i\nu}p_{i\nu}^n
(1-ap_{i\nu})^{-k}(1-A/p_{i\nu})^{-K}(1-bp_{i\nu})^{-l}(1-B/p_{i\nu})^{-L}, \\
&&\psi_j^{(n)}(k,K,l,L)=\sum_{\mu=1}^2b_{j\mu}q_{j\mu}^n
(1+aq_{j\mu})^k(1+A/q_{j\mu})^K(1+bq_{j\mu})^l(1+B/q_{j\mu})^L,
\end{eqnarray*}
where $p_{i\nu}$, $q_{j\mu}$, $a_{i\nu}$, $b_{j\mu}$ are arbitrary constants. Obviously, the $\phi_i^{(n)}$ and $\psi_j^{(n)}$ defined satisfy the above linear dispersion relations, therefore the determinant with above defined element  $m_{ij}^{(n)}(k,K,l,L)$ satisfy the bilinear equations (\ref{bilin1}).

Next, we proceed to the reductions. By imposing 
$$
p_{i2}=-\frac{A}{a}\frac{1-ap_{i1}}{p_{i1}-A},\quad
q_{j2}=-\frac{A}{a}\frac{1+aq_{j1}}{q_{j1}+A},
$$
we can show that $\tau_n$ satisfies the following reduction condition,
\begin{equation*}
\tau_n(k+1,K-1,l,L)=\tau_n(k,K,l,L)\prod_{i=1}^N
\frac{q_{i1}q_{i2}}{p_{i1}p_{i2}},
\end{equation*}
it follows that the bilinear equations (\ref{bilin1}) are reduced to
\begin{equation}
\left\{
\begin{array}{l}
(1-aA)\sigma_{k+1}^t(n)\sigma_{k-1}^t(n)-\sigma_k^t(n)\sigma_k^t(n)
+aA\sigma_k^t(n+1)\sigma_k^t(n-1)=0, \\[5pt]
A(a-b)(1-aB)\sigma_k^{t+1}(n+1)\sigma_k^t(n)
-a(A-B)(1-Ab)\sigma_k^t(n+1)\sigma_k^{t+1}(n) \\[5pt]
\quad +Ab(1-aA)\sigma_{k-1}^{t+1}(n+1)\sigma_{k+1}^t(n)
-aB(1-aA)\sigma_{k+1}^t(n+1)\sigma_{k-1}^{t+1}(n)=0,
\end{array}
\right.
\label{bilin2}
\end{equation}
by taking $\sigma_k^t(n)=\tau_n(k,0,t,t)$.
Furthermore by taking
$$
A=-\epsilon\bar a,\quad
B=-\epsilon\bar b,\quad
q_{j1}=\frac{\bar a}{a}\frac{\bar p_{j1}+\epsilon a}{1-\bar a\bar p_{j1}},\quad
b_{j1}=\frac{\bar a_{j1}}{a(1-\bar a\bar p_{j1})},\quad
b_{j2}=\frac{\bar a_{j2}}{a(1-\bar a\bar p_{j2})},
$$
where $\epsilon=+1$ or $-1$,
$\sigma_k^t(n)$ satisfies the complex conjugate condition,
$$
\sigma_k^t(-n)G=\overline{\sigma_k^t(n)G},
$$
with a gauge factor $G$ which is given below in eq.(\ref{gauge}).
In order to simplify the final expression, let us parametrise by
$$
|a|^2=\epsilon\frac{r-2+1/r}{4},\quad
b=\frac{2a}{1-r}\frac{1-rc}{1+c},\quad
p_{i1}=\frac{1-1/r}{2a}\frac{1-rp_i}{1+p_i},
$$
$$
a_{i1}=\epsilon\frac{r-1/r}{4}a_i,\quad
a_{i2}=-a(\bar a+\epsilon p_{i2})b_i,\quad
$$
where $r$ is a real constant, $c$ is a complex constant,
$p_i$ represents the complex wave number of the $i$th breather,
$a_i$ and $b_i$ stand for the complex phase constants of the $i$th breather.
We comment that $\epsilon$ is equal to the sign of $r$
because of the positivity of $|a|^2$.
Then we have
\begin{eqnarray*}
&&\tilde m_{ij}^n(k,K,l,L) \\
&&:=m_{ij}^n(k,K,l,L)\frac{-4}{(1-r)(1+p_i)(1+\bar p_j)}
\left(-r\frac{1-\bar p_j^2}{1-r^2\bar p_j^2}\right)^n
r^{K-k}(rc)^{-l}
\left(r\bar c\frac{1-p_i^2/\bar c^2}{1-r^2p_i^2}
\frac{1-\bar p_j^2}{1-r^2\bar c^2\bar p_j^2}\right)^L \\
&&=\frac{a_i\bar a_j}{1-rp_i\bar p_j}
\left(\frac{1-rp_i}{1+p_i}\frac{1+\bar p_j}{1-r\bar p_j}\right)^n
\left(\frac{1+p_i}{1+rp_i}\frac{1+\bar p_j}{1+r\bar p_j}\right)^k
\left(\frac{1-rp_i}{1-p_i}\frac{1-r\bar p_j}{1-\bar p_j}\right)^K \\
&&\hskip50pt\times
\left(\frac{1+p_i}{1+rcp_i}\frac{1+\bar p_j/c}{1+r\bar p_j}\right)^l
\left(\frac{1+p_i/\bar c}{1+rp_i}\frac{1+\bar p_j}{1+r\bar c\bar p_j}\right)^L
\\
&&-\frac{a_i\bar b_j}{1+rp_i\bar p_j}
\left(\frac{1-rp_i}{1+p_i}\frac{1-\bar p_j}{1+r\bar p_j}\right)^n
\left(\frac{1+p_i}{1+rp_i}\frac{1-\bar p_j}{1-r\bar p_j}\right)^k
\left(\frac{1-rp_i}{1-p_i}\frac{1+r\bar p_j}{1+\bar p_j}\right)^K \\
&&\hskip50pt\times
\left(\frac{1+p_i}{1+rcp_i}\frac{1-\bar p_j/c}{1-r\bar p_j}\right)^l
\left(\frac{1+p_i/\bar c}{1+rp_i}\frac{1-\bar p_j}{1-r\bar c\bar p_j}\right)^L
\\
&&-\frac{b_i\bar a_j}{1+rp_i\bar p_j}
\left(\frac{1+rp_i}{1-p_i}\frac{1+\bar p_j}{1-r\bar p_j}\right)^n
\left(\frac{1-p_i}{1-rp_i}\frac{1+\bar p_j}{1+r\bar p_j}\right)^k
\left(\frac{1+rp_i}{1+p_i}\frac{1-r\bar p_j}{1-\bar p_j}\right)^K \\
&&\hskip50pt\times
\left(\frac{1-p_i}{1-rcp_i}\frac{1+\bar p_j/c}{1+r\bar p_j}\right)^l
\left(\frac{1-p_i/\bar c}{1-rp_i}\frac{1+\bar p_j}{1+r\bar c\bar p_j}\right)^L
\\
&&+\frac{b_i\bar b_j}{1-rp_i\bar p_j}
\left(\frac{1+rp_i}{1-p_i}\frac{1-\bar p_j}{1+r\bar p_j}\right)^n
\left(\frac{1-p_i}{1-rp_i}\frac{1-\bar p_j}{1-r\bar p_j}\right)^k
\left(\frac{1+rp_i}{1+p_i}\frac{1+r\bar p_j}{1+\bar p_j}\right)^K \\
&&\hskip50pt\times
\left(\frac{1-p_i}{1-rcp_i}\frac{1-\bar p_j/c}{1-r\bar p_j}\right)^l
\left(\frac{1-p_i/\bar c}{1-rp_i}\frac{1-\bar p_j}{1-r\bar c\bar p_j}\right)^L,
\end{eqnarray*}
and the bilinear equations (\ref{bilin2}) are reduced to
\begin{equation}
\left\{
\begin{array}{l}
(1+r)^2\tau_{k+1}^t(n)\tau_{k-1}^t(n)-4r\tau_k^t(n)\tau_k^t(n)
-(1-r)^2\tau_k^t(n+1)\tau_k^t(n-1)=0, \\[5pt]
(1-c)(1+r\bar c)\tau_k^{t+1}(n+1)\tau_k^t(n)
-(1-\bar c)(1+rc)\tau_k^t(n+1)\tau_k^{t+1}(n) \\[5pt]
\quad -(1+\bar c)(1-rc)\tau_{k-1}^{t+1}(n+1)\tau_{k+1}^t(n)
+(1+c)(1-r\bar c)\tau_{k+1}^t(n+1)\tau_{k-1}^{t+1}(n)=0,
\end{array}
\right.
\end{equation}
where $\tau_k^t(n)$ is given by
$$
\tau_k^t(n)=\det_{1\le i,j\le N}\left(\tilde m_{ij}^{n}(k,0,t,t)\right)
=\sigma_k^t(n)G,
$$
\begin{equation}
G=\prod_{i=1}^N\frac{-4}{(1-r)(1+p_i)(1+\bar p_i)}
\left(-r\frac{1-\bar p_i^2}{1-r^2\bar p_i^2}\right)^n
r^{-k}
\left(\frac{\bar c}{c}\frac{1-p_i^2/\bar c^2}{1-r^2p_i^2}
\frac{1-\bar p_i^2}{1-r^2\bar c^2\bar p_i^2}\right)^t,
\label{gauge}
\end{equation}
which is nothing but $\tau_k^t(n)$ in Theorem 1.
For $f_k^t=\tau_k^t(0)$, $g_k^t=\tau_k^t(1)$, $\bar g_k^t=\tau_k^t(-1)$,
the bilinear equations are written as
\begin{equation*}
\left\{
\begin{array}{l}
(1+r)^2f_{k+1}^tf_{k-1}^t-4rf_k^tf_k^t-(1-r)^2g_k^t\bar g_k^t=0, \\[5pt]
(1-c)(1+r\bar c)g_k^{t+1}f_k^t
-(1-\bar c)(1+rc)g_k^tf_k^{t+1} \\[5pt]
\quad -(1+\bar c)(1-rc)g_{k-1}^{t+1}f_{k+1}^t
+(1+c)(1-r\bar c)g_{k+1}^tf_{k-1}^{t+1}=0,
\end{array}
\right.
\end{equation*}
from which the fully discrete NLS eq.(\ref{dNLS}) is straightforwardly derived
through the variable transformation (\ref{vartr}).
This completes the proof of Theorem 1.
\end{proof}
%\hfill\rule{7pt}{10pt}

In the case of $N=1$, we have the 1-breather solution,
\begin{eqnarray*}
&&\tau_k^t(n)=\frac{1}{1-r|p|^2}
\left(|a|^2\varphi_n(p)\overline{\varphi_{-n}(p)}
+|b|^2\varphi_n(-p)\overline{\varphi_{-n}(-p)}\right) \\
&&\qquad -\frac{1}{1+r|p|^2}
\left(a\bar b\varphi_n(p)\overline{\varphi_{-n}(-p)}
+\bar ab\varphi_n(-p)\overline{\varphi_{-n}(p)}\right),
\end{eqnarray*}
with (\ref{vartr})-(\ref{varphi}) where $p$, $a$, $b$ are complex constants
(the index $1$ of $p_1$, $a_1$, $b_1$ are omitted for notational simplicity).
The Akhmediev breather which is the breather solution localized in time $t$
can be derived by taking the wave number $p$ pure imaginary.
If we require the regularity of the solution only on the lattice points,
i.e., $f_k^t\ne 0$ for integers $k$ and $t$,
then $r$ can be either positive or negative and the discrete NLS equations
(\ref{dNLS}) of both focusing type and defocusing type admit regular breather
solutions by locating the singularities (if exist) off the lattice points.
On the other hand if we require the solution to be regular in whole real space,
i.e., $f_k^t$ is non-zero for all real numbers $k$ and $t$,
then $r$ has to be positive and only the focusing discrete NLS admits
the regular breathers for generic parameters.
Some exceptional regular solutions for $r<0$ can be derived from the above
$\tau_k^t(n)$ but usually they are not called breathers and we don't discuss
about such solutions.
\begin{figure}[htp]
	\centering
		{
		\includegraphics{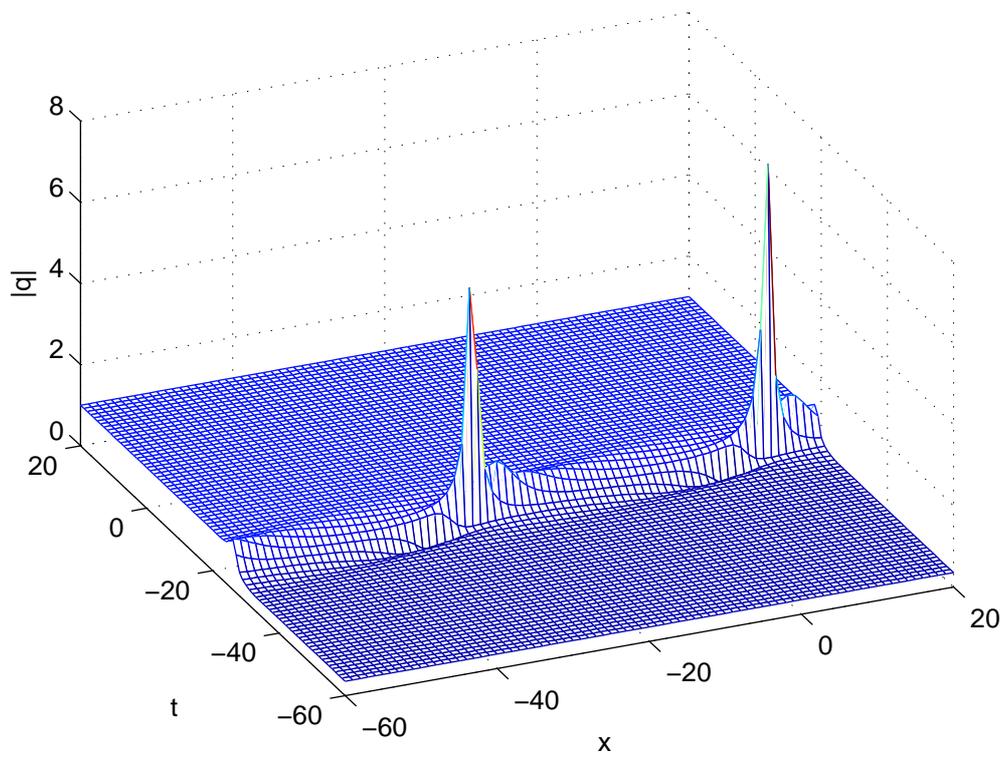}
			}
	\caption{One-breather solution with parameters $r=2.0$, $c=3+2\rm{i}$, $p=0.5\rm{i}$}
	\label{1-br} 
	\end{figure}

\noindent
\section{Rogue wave solution for fully discrete NLS equation}

The rogue wave solution of rational function type can be derived
as a limit of the breather solution.
In Theorem 1, we take
$$
a_i=\frac{1}{2}\Big(1+\sum_{\nu=1}^ic_{\nu}p_i^{2\nu-1}\Big), \quad
b_i=\frac{1}{2}\Big(1-\sum_{\nu=1}^ic_{\nu}p_i^{2\nu-1}\Big),
$$
scale the $\tau$ function by $\displaystyle
\tau_k^t(n)/\prod_{i=1}^N(p_i\bar p_i)^{2i-1}$
and finally take the limit $p_i\to0$ successively for $i=1,2,\ldots,N$.
Then the leading order of $\tau_k^t(n)$ in $p_i$'s gives a polynomial
of $k$ and $t$ which turns to be the rogue wave solution.
This result is summarized in the following theorem.

\begin{theorem}
The $N$th order rogue wave solution for the fully discrete NLS equation
(\ref{dNLS}) is given by
$$
\tau_k^t(n)=\det_{1\le i,j\le N}\left(B_{ij}^{(n)}(k,t)\right),
\qquad
B_{ij}^{(n)}=\sum_{\nu=1}^{2\min(i,j)}
r^{\nu-1}\Phi_{2i-\nu}^{(n)}\overline{\Phi_{2j-\nu}^{(-n)}},
$$
$$
\Phi_i^{(n)}=S_i(x(n))
+\sum_{\mu=1}^{\left[\frac{i+1}{2}\right]}c_{\mu}S_{i+1-2\mu}(x(n)),
\qquad
x(n)=(x_1(n),x_2(n),\ldots,x_h(n),\ldots),
$$
$$
x_h(n)=\frac{(-1)^h}{h}
\left((1-(-r)^h)n-(1-r^h)k-(1-r^h+(1/\bar c)^h-(rc)^h)t\right),
\quad\hbox{for }h=1,2,\ldots,
$$
with the variable transformations (\ref{vartr})-(\ref{fg}),
where $[\ ]$ means the Gauss symbol and $S_{\mu}(x)$ is the so-called
elementary Schur function defined by $\displaystyle
\sum_{\mu=0}^{\infty}S_{\mu}(x)\lambda^{\mu}
=\exp\sum_{h=1}^{\infty}x_h\lambda^h$.
This solution has $N$ complex parameters $c_\mu$, $\mu=1,2,\ldots,N$.
\end{theorem}
\noindent
\begin{proof}
Firstly $\varphi_n(p)$ in (\ref{varphi}) is written as
$\displaystyle\varphi_n(p)=\sum_{\mu=0}^{\infty}S_{\mu}(x(n))p^{\mu}$.
We rewrite $A_{ij}^{(n)}$ in Theorem 1 as
\begin{eqnarray*}
&&A_{ij}^{(n)}=\frac{1}{1-(rp_i\bar p_j)^2}
\Big(a_i\varphi_n(p_i)-b_i\varphi_n(-p_i)\Big)
\Big(\overline{a_j\varphi_{-n}(p_j)-b_j\varphi_{-n}(-p_j)}\Big)  \\
&&\qquad +\frac{rp_i\bar p_j}{1-(rp_i\bar p_j)^2}
\Big(a_i\varphi_n(p_i)+b_i\varphi_n(-p_i)\Big)
\Big(\overline{a_j\varphi_{-n}(p_j)+b_j\varphi_{-n}(-p_j)}\Big).
\end{eqnarray*}
Denoting $a_i-b_i=p_id_i$ and $a_i+b_i=s_i$,
the four factors in the above expression are written as
\begin{eqnarray*}
&&a_i\varphi_n(p_i)-b_i\varphi_n(-p_i)
=\sum_{\mu=0}^{\infty}S_{\mu}(x(n))p_i^{\mu}(a_i-(-1)^{\mu}b_i) \\
&&\qquad =(p_id_i,p_is_i,p_i^3d_i,p_i^3s_i,\cdots)
\,{}^t(S_0(x(n)),S_1(x(n)),S_2(x(n)),S_3(x(n)),\cdots), \\
&&\overline{a_j\varphi_{-n}(p_j)-b_j\varphi_{-n}(-p_j)}
=\sum_{\mu=0}^{\infty}\overline{S_{\mu}(x(-n))p_j^{\mu}(a_j-(-1)^{\mu}b_j)} \\
&&\qquad
=(S_0(\overline{x(-n)}),S_1(\overline{x(-n)}),
S_2(\overline{x(-n)}),S_3(\overline{x(-n)}),\cdots)
\,{}^t(\bar p_j\bar d_j,\bar p_j\bar s_j,
\bar p_j^3\bar d_j,\bar p_j^3\bar s_j,\cdots), \\
&&a_i\varphi_n(p_i)+b_i\varphi_n(-p_i)
=\sum_{\mu=0}^{\infty}S_{\mu}(x(n))p_i^{\mu}(a_i+(-1)^{\mu}b_i) \\
&&\qquad =(s_i,p_i^2d_i,p_i^2s_i,p_i^4d_i,\cdots)
\,{}^t(S_0(x(n)),S_1(x(n)),S_2(x(n)),S_3(x(n)),\cdots), \\
&&\overline{a_j\varphi_{-n}(p_j)+b_j\varphi_{-n}(-p_j)}
=\sum_{\mu=0}^{\infty}\overline{S_{\mu}(x(-n))p_j^{\mu}(a_j+(-1)^{\mu}b_j)} \\
&&\qquad
=(S_0(\overline{x(-n)}),S_1(\overline{x(-n)}),
S_2(\overline{x(-n)}),S_3(\overline{x(-n)}),\cdots)
\,{}^t(\bar s_j,\bar p_j^2\bar d_j,\bar p_j^2\bar s_j,
\bar p_j^4\bar d_j,\cdots),
\end{eqnarray*}
where ${}^tv$ means transpose of $v$.
Thus we obtain
\begin{eqnarray*}
&&A_{ij}^{(n)}=\sum_{\lambda=0}^{\infty}(rp_i\bar p_j)^{2\lambda}
(p_id_i,p_is_i,p_i^3d_i,p_i^3s_i,\cdots)
\,{}^t(S_0(x(n)),S_1(x(n)),S_2(x(n)),S_3(x(n)),\cdots) \\
&&\hskip40pt\times
(S_0(\overline{x(-n)}),S_1(\overline{x(-n)}),
S_2(\overline{x(-n)}),S_3(\overline{x(-n)}),\cdots)
\,{}^t(\bar p_j\bar d_j,\bar p_j\bar s_j,
\bar p_j^3\bar d_j,\bar p_j^3\bar s_j,\cdots) \\
&&\qquad +\sum_{\lambda=0}^{\infty}(rp_i\bar p_j)^{2\lambda+1}
(s_i,p_i^2d_i,p_i^2s_i,p_i^4d_i,\cdots)
\,{}^t(S_0(x(n)),S_1(x(n)),S_2(x(n)),S_3(x(n)),\cdots) \\
&&\hskip40pt\times
(S_0(\overline{x(-n)}),S_1(\overline{x(-n)}),
S_2(\overline{x(-n)}),S_3(\overline{x(-n)}),\cdots)
\,{}^t(\bar s_j,\bar p_j^2\bar d_j,\bar p_j^2\bar s_j,
\bar p_j^4\bar d_j,\cdots) \\
&&\qquad
=\begin{pmatrix}
p_id_i&p_is_i&p_i^3d_i&p_i^3s_i&\cdots
\end{pmatrix}
\raisebox{-30pt}{$\begin{pmatrix}
S_0(x(n))&&&&\raisebox{-5pt}{\smash{\Large 0}} \\
S_1(x(n))&S_0(x(n)) \\
S_2(x(n))&S_1(x(n))&S_0(x(n)) \\
S_3(x(n))&S_2(x(n))&S_1(x(n))&S_0(x(n)) \\
\vdots&\vdots&\vdots&&\ddots
\end{pmatrix}$} \\
&&\qquad\times
\begin{pmatrix}
1&&&&\raisebox{-5pt}{\smash{\Large 0}} \\
&r \\
&&r^2 \\
&&&r^3 \\
\mbox{\smash{\Large 0}}&&&&\ddots
\end{pmatrix}
\begin{pmatrix}
S_0(\overline{x(-n)})&S_1(\overline{x(-n)})
&S_2(\overline{x(-n)})&S_3(\overline{x(-n)})&\cdots \\
&S_0(\overline{x(-n)})&S_1(\overline{x(-n)})&S_2(\overline{x(-n)})&\cdots \\
&&S_0(\overline{x(-n)})&S_1(\overline{x(-n)})&\cdots \\
&&&S_0(\overline{x(-n)}) \\
\mbox{\smash{\Large 0}}&&&&\ddots
\end{pmatrix}
\begin{pmatrix}
\bar p_j\bar d_j \\
\bar p_j\bar s_j \\
\bar p_j^3\bar d_j \\
\bar p_j^3\bar s_j \\
\vdots
\end{pmatrix}.
\end{eqnarray*}
Therefore $\tau_k^t(n)$ in Theorem 1 is given in the form of
the following determinant,
\begin{eqnarray*}
&&\tau_k^t(n)=\left|
\raisebox{8pt}{$\begin{pmatrix}
p_1d_1&p_1s_1&p_1^3d_1&p_1^3s_1&\cdots \\
p_2d_2&p_2s_2&p_2^3d_2&p_2^3s_2&\cdots \\
\vdots&\vdots&\vdots&\vdots \\
p_Nd_N&p_Ns_N&p_N^3d_N&p_N^3s_N&\cdots
\end{pmatrix}$}
\begin{pmatrix}
S_0(x(n))&&&&\raisebox{-5pt}{\smash{\Large 0}} \\
S_1(x(n))&S_0(x(n)) \\
S_2(x(n))&S_1(x(n))&S_0(x(n)) \\
S_3(x(n))&S_2(x(n))&S_1(x(n))&S_0(x(n)) \\
\vdots&\vdots&\vdots&&\ddots
\end{pmatrix}\right. \\
&&\hskip40pt\times
\begin{pmatrix}
1&&&&\raisebox{-5pt}{\smash{\Large 0}} \\
&r \\
&&r^2 \\
&&&r^3 \\
\mbox{\smash{\Large 0}}&&&&\ddots
\end{pmatrix}
\begin{pmatrix}
S_0(\overline{x(-n)})&S_1(\overline{x(-n)})
&S_2(\overline{x(-n)})&S_3(\overline{x(-n)})&\cdots \\
&S_0(\overline{x(-n)})&S_1(\overline{x(-n)})&S_2(\overline{x(-n)})&\cdots \\
&&S_0(\overline{x(-n)})&S_1(\overline{x(-n)})&\cdots \\
&&&S_0(\overline{x(-n)}) \\
\mbox{\smash{\Large 0}}&&&&\ddots
\end{pmatrix} \\
&&\hskip40pt\times
\left.\begin{pmatrix}
\bar p_1\bar d_1&\bar p_2\bar d_2&\cdots&\bar p_N\bar d_N \\
\bar p_1\bar s_1&\bar p_2\bar s_2&\cdots&\bar p_N\bar s_N \\
\bar p_1^3\bar d_1&\bar p_2^3\bar d_2&\cdots&\bar p_N^3\bar d_N \\
\bar p_1^3\bar s_1&\bar p_2^3\bar s_2&\cdots&\bar p_N^3\bar s_N \\
\vdots&\vdots&&\vdots \\
\end{pmatrix}\right|.
\end{eqnarray*}

Now let us take
$$
a_i=\frac{1}{2}\Big(1+\sum_{\nu=1}^ic_{\nu}p_i^{2\nu-1}\Big), \quad
b_i=\frac{1}{2}\Big(1-\sum_{\nu=1}^ic_{\nu}p_i^{2\nu-1}\Big).
$$
Then we have $\displaystyle d_i=\sum_{\nu=1}^ic_{\nu}p_i^{2\nu-2}$ and $s_i=1$,
and the above $\tau_k^t(n)$ is $O(p_1\bar p_1p_2\bar p_2\cdots p_N\bar p_N)$
as $p_i\to 0$ for $1\le i\le N$.
In order to take the lowest order in $p_1$, we consider the limit,
$\displaystyle\tilde\tau_k^t(n):=\lim_{p_1\to 0}\tau_k^t(n)/(p_1\bar p_1)$.
In this limit, the leading order becomes
$O((p_2\bar p_2\cdots p_N\bar p_N)^3)$,
thus for picking up the lowest order in $p_2$, we take the limit,
$\displaystyle\lim_{p_2\to 0}\tilde\tau_k^t(n)/(p_2\bar p_2)^3$.
So the leading order becomes $O((p_3\bar p_3\cdots p_N\bar p_N)^5)$.
Repeating this procedure, finally we obtain the $\tau$ function
of rogue wave solution from that of breather, $\tau_k^t(n)$,
\begin{eqnarray*}
&&\lim_{p_N\to 0}\cdots\lim_{p_2\to 0}\lim_{p_1\to 0}
\frac{\tau_k^t(n)}
{p_1\bar p_1p_2^3\bar p_2^3\cdots p_N^{2N-1}\bar p_N^{2N-1}} \\
&&=\left|
\begin{pmatrix}
c_1&1&&&&&&&&\raisebox{-5pt}{\smash{\Large 0}} \\
c_2&0&c_1&1 \\
c_3&0&c_2&0&c_1&1 \\
\vdots&\vdots&\vdots&\vdots&&&\ddots \\
c_N&0&c_{N-1}&0&\cdots&\cdots&\cdots&c_1&1&
\end{pmatrix}
\begin{pmatrix}
S_0(x(n))&&&&\raisebox{-5pt}{\smash{\Large 0}} \\
S_1(x(n))&S_0(x(n)) \\
S_2(x(n))&S_1(x(n))&S_0(x(n)) \\
S_3(x(n))&S_2(x(n))&S_1(x(n))&S_0(x(n)) \\
\vdots&\vdots&\vdots&&\ddots
\end{pmatrix}\right. \\
&&\hskip30pt\times
\begin{pmatrix}
1&&&&\raisebox{-5pt}{\smash{\Large 0}} \\
&r \\
&&r^2 \\
&&&r^3 \\
\mbox{\smash{\Large 0}}&&&&\ddots
\end{pmatrix}
\begin{pmatrix}
S_0(\overline{x(-n)})&S_1(\overline{x(-n)})
&S_2(\overline{x(-n)})&S_3(\overline{x(-n)})&\cdots \\
&S_0(\overline{x(-n)})&S_1(\overline{x(-n)})&S_2(\overline{x(-n)})&\cdots \\
&&S_0(\overline{x(-n)})&S_1(\overline{x(-n)})&\cdots \\
&&&S_0(\overline{x(-n)}) \\
\mbox{\smash{\Large 0}}&&&&\ddots
\end{pmatrix} \\
&&\hskip30pt\times
\left.\begin{pmatrix}
\bar c_1&\bar c_2&\bar c_3&\cdots&\bar c_N \\
1&0&0&\cdots&0 \\
&\bar c_1&\bar c_2&\cdots&\bar c_{N-1} \\
&1&0&\cdots&0 \\
&&\bar c_1&&\vdots \\
&&1&&\vdots \\
&&&\ddots&\vdots \\
&&&&\bar c_1 \\
&&&&1 \\
\mbox{\smash{\Large 0}}
\end{pmatrix}\right| \\
&&=\left|
\begin{pmatrix}
\Phi_1^{(n)}&\Phi_0^{(n)}&&&&&&&&\raisebox{-5pt}{\smash{\Large 0}} \\
\Phi_3^{(n)}&\Phi_2^{(n)}&\Phi_1^{(n)}&\Phi_0^{(n)} \\
\Phi_5^{(n)}&\Phi_4^{(n)}&\Phi_3^{(n)}&\Phi_2^{(n)}&\Phi_1^{(n)}&\Phi_0^{(n)}  \\
\vdots&\vdots&\vdots&\vdots&&&\ddots \\
\Phi_{2N-1}^{(n)}&\Phi_{2N-2}^{(n)}&\Phi_{2N-3}^{(n)}&\Phi_{2N-4}^{(n)}
&\cdots&\cdots&\cdots&\Phi_1^{(n)}&\Phi_0^{(n)}
\end{pmatrix}\right.
\begin{pmatrix}
1&&&&\raisebox{-5pt}{\smash{\Large 0}} \\
&r \\
&&r^2 \\
&&&r^3 \\
\mbox{\smash{\Large 0}}&&&&\ddots
\end{pmatrix} \\
&&\hskip30pt\times
\left.\begin{pmatrix}
\overline{\Phi_1^{(-n)}}&\overline{\Phi_3^{(-n)}}
&\overline{\Phi_5^{(-n)}}&\cdots&\overline{\Phi_{2N-1}^{(-n)}} \\
\overline{\Phi_0^{(-n)}}&\overline{\Phi_2^{(-n)}}
&\overline{\Phi_4^{(-n)}}&\cdots&\overline{\Phi_{2N-2}^{(-n)}} \\
&\overline{\Phi_1^{(-n)}}&\overline{\Phi_3^{(-n)}}
&\cdots&\overline{\Phi_{2N-3}^{(-n)}} \\
&\overline{\Phi_0^{(-n)}}&\overline{\Phi_2^{(-n)}}
&\cdots&\overline{\Phi_{2N-4}^{(-n)}} \\
&&\overline{\Phi_1^{(-n)}}&&\vdots \\
&&\overline{\Phi_0^{(-n)}}&&\vdots \\
&&&\ddots&\vdots \\
&&&&\overline{\Phi_1^{(-n)}} \\
&&&&\overline{\Phi_0^{(-n)}} \\
\mbox{\smash{\Large 0}}
\end{pmatrix}\right|.
\end{eqnarray*}
By calculating the matrix element, it is easy to see that the above determinant
is equal to $\displaystyle\det_{1\le i,j\le N}\left(B_{ij}^{(n)}(k,t)\right)$.
We completed the proof of Theorem 2.
\end{proof}
%\hfill\rule{7pt}{10pt}

\begin{figure}[htp]
	\centering
		\subfigure[]
	{
		\includegraphics[width=2.2in]{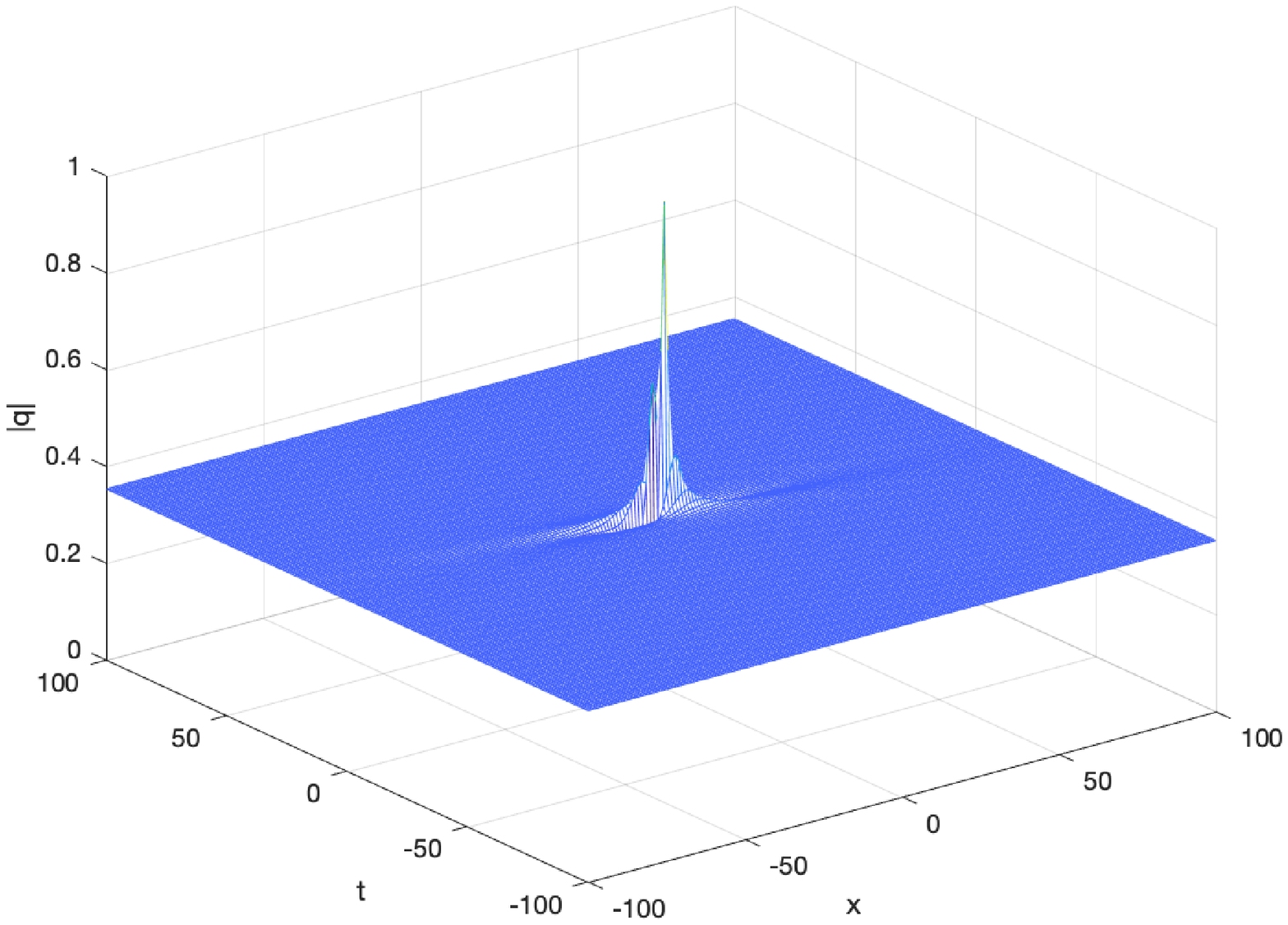}
			}
	\hspace*{3em}
			\subfigure[]
		{
		\includegraphics[width=2.2in]{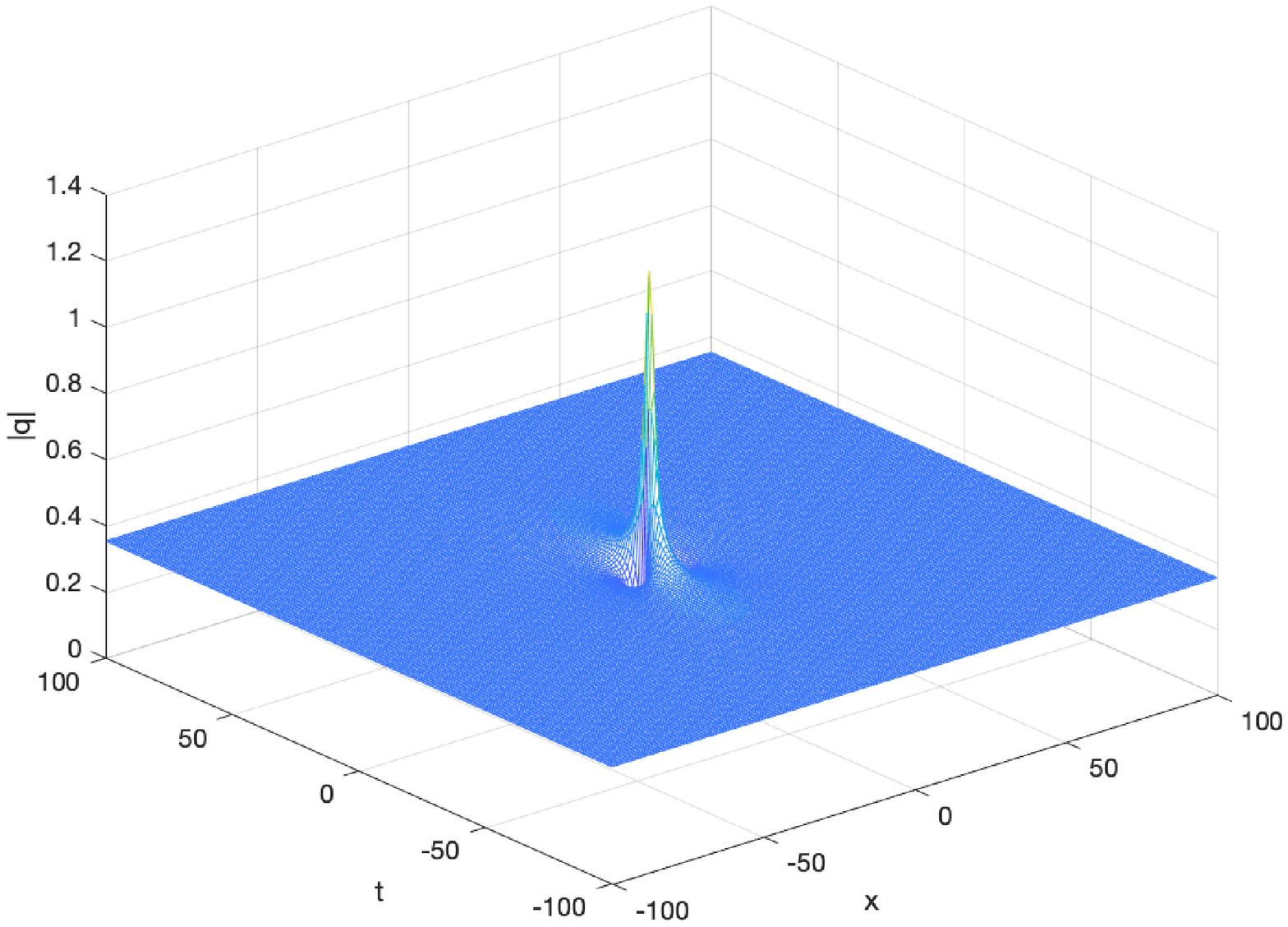}
			}
	\caption{First order rogue wave solution (a) $r=2.0$, $c=1+2\rm{i}$, $c_1=2$ (b) $r=0.5$, $c=1+2\rm{i}$, $c_1=2$.}
	\label{1st_rw}
\end{figure}

\begin{figure}[htp]
	\centering
		\subfigure[]
	{
		\includegraphics[width=2.2in]{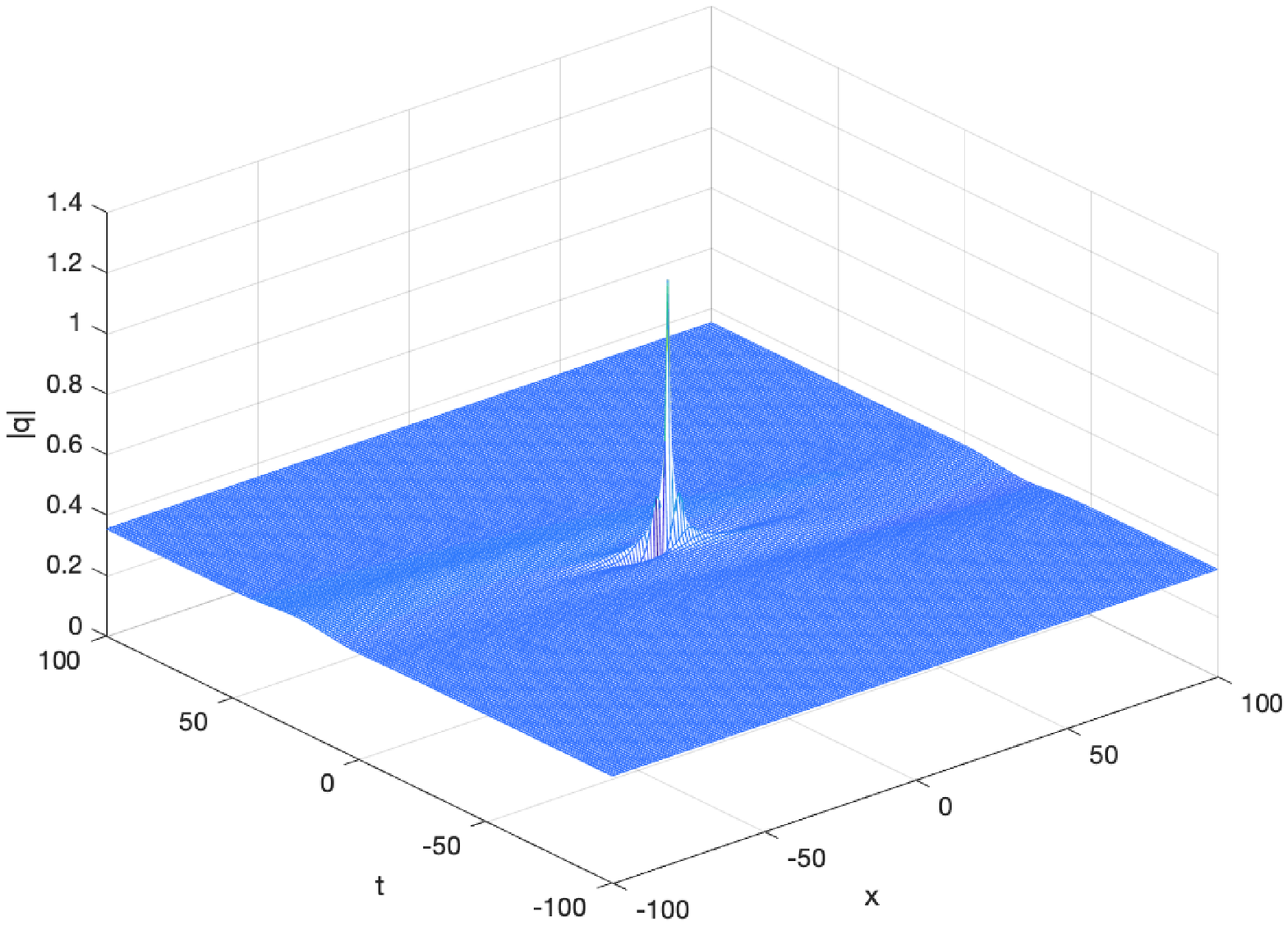}
			}
	\hspace*{3em}
		\subfigure[]
		{
		\includegraphics[width=2.2in]{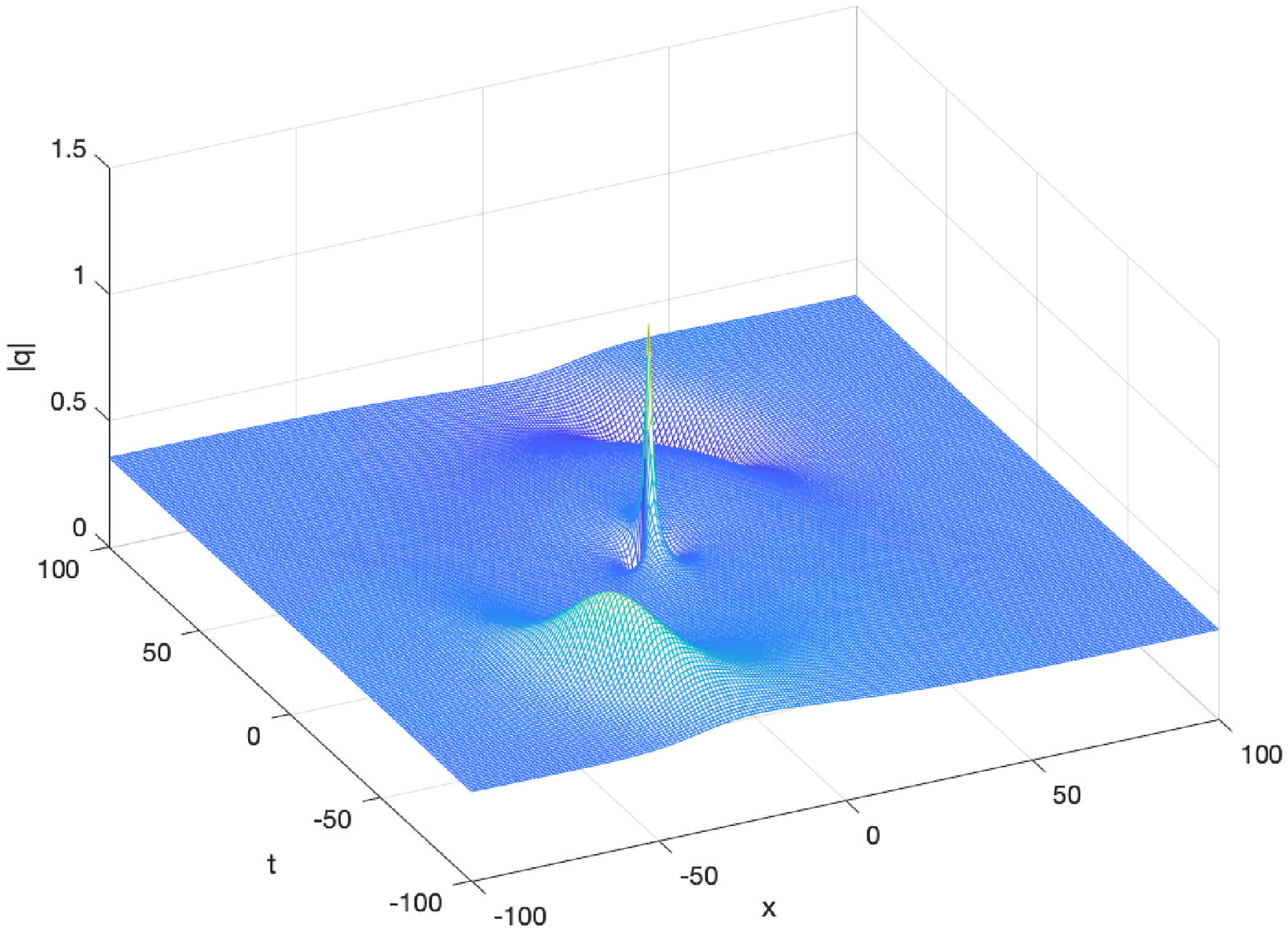}
			}
	\caption{Second order rogue wave solution (a) $r=2.0$, $c=1+2\rm{i}$, $c_1=2.0$, $c_2=2+2\rm{i}$; (b)  $r=0.5$, $c=2+\rm{i}$, $c_1=2$, $c_2=2+2\rm{i}$. }
	\label{2nd_rw}
	\end{figure}

By taking $N=1$ we obtain the fully discrete Peregrine rogue wave solution,
\begin{eqnarray}
&&\tau_k^t(n)=\left(-(1+r)n+(1-r)k+(1-r+1/\bar c-rc)t+c_1\right) \nonumber\\
&&\hskip30pt\times\left((1+r)n+(1-r)k+(1-r+1/c-r\bar c)t+\bar c_1\right)+r,
\label{Peregrine}
\end{eqnarray}
where $c_1$ is a complex constant.
Similarly to the breather solution, there are rogue wave solutions
regular on the lattice for both focusing case ($r>0$) and defocusing
case ($r<0$), since if there are zeros of $f_k^t$ we can avoid
explosion of solution by displacing the zeros off the lattice points.
However for regularity of the solution on the real two dimensional space
of $(k,t)$, we have to take $r$ positive. First order and second-order rogue wave solutions are shown in  Figs. \ref{1st_rw} and \ref{2nd_rw}, respectively. 

\begin{figure}[htp]
	\centering
		\subfigure[]
	{
		\includegraphics[width=2.2in]{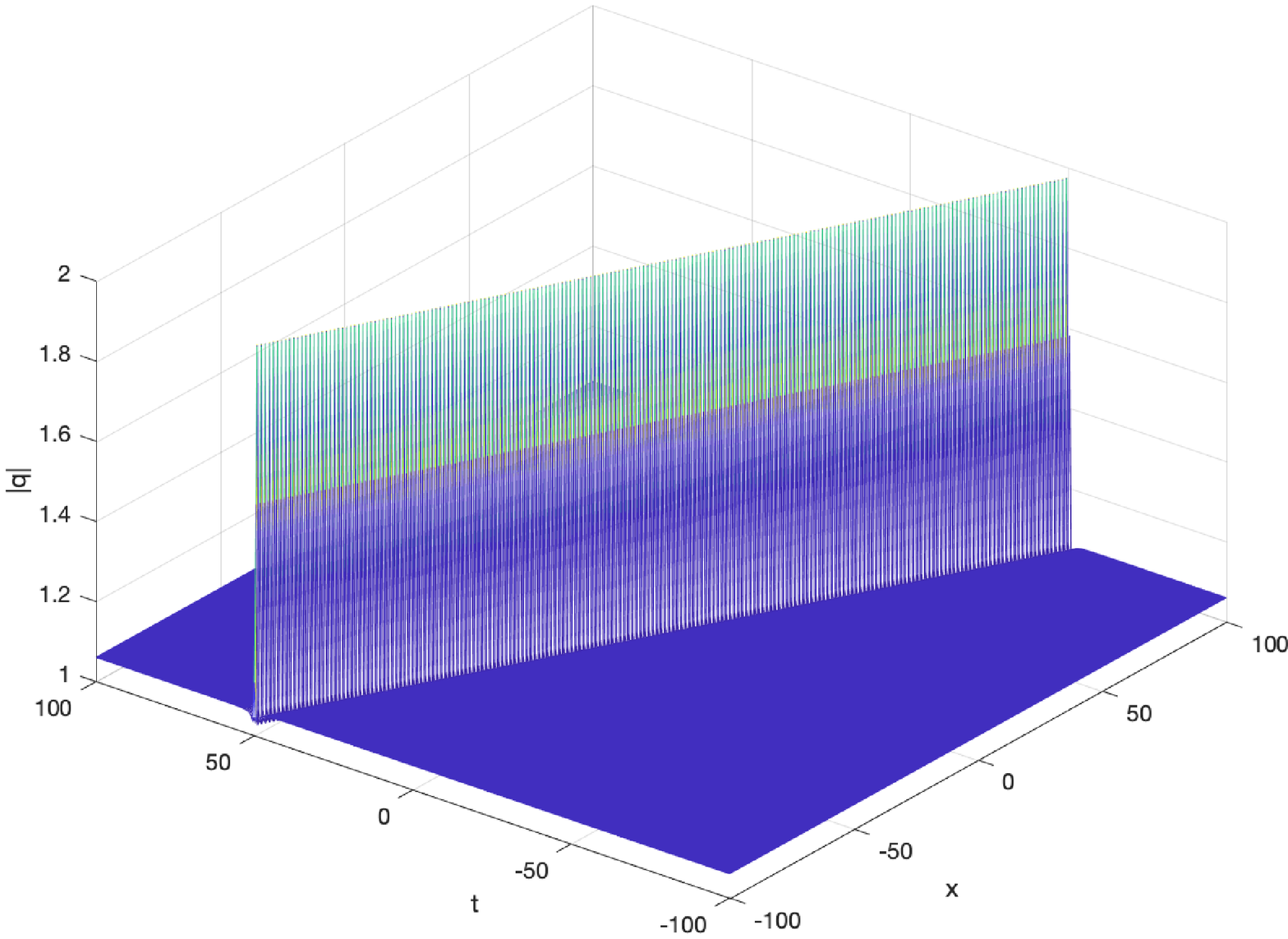}
			}
	\hspace*{3em}
		\subfigure[]
		{
		\includegraphics[width=2.2in]{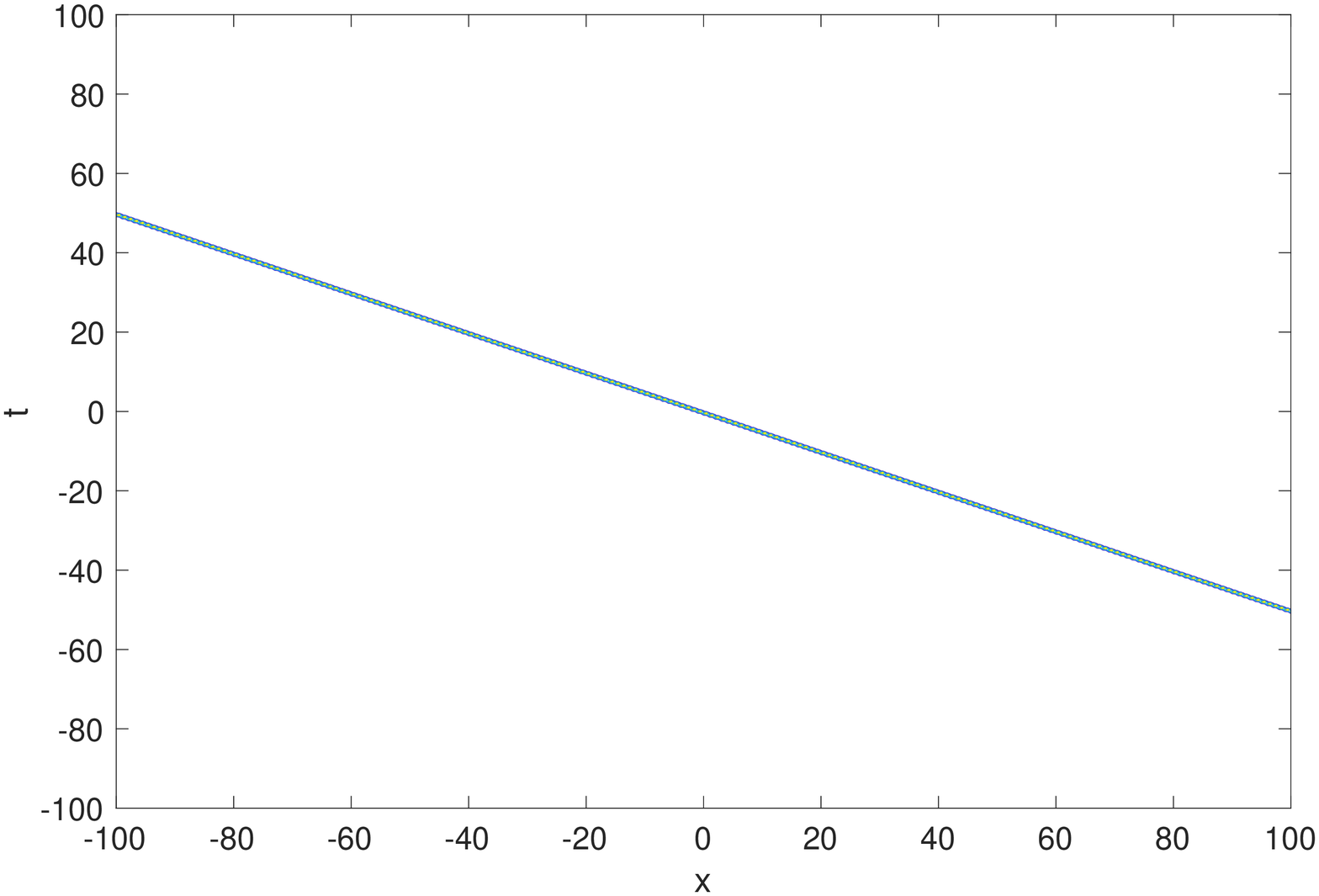}
			}
	\caption{Traveling wave solution with $r=-0.5$, $c=1.0$, $c_1=1+\rm{i}$: (a) profile, (b) contour plot.} 
	\label{tw}
\end{figure}

There is an exceptional regular solution for $r<0$ which is obtained
by taking $c$ real and $({\rm Im}\,c_1)^2>-r$ in (\ref{Peregrine}),
but this is not a rogue wave solution but a traveling wave solution. An example is shown in Fig. \ref{tw}.

\section{Concluding Remarks}
Even though the study of rogue waves has attracted much attention in more than one decade, the rogue wave solution in fully discrete integrable systems has not been reported yet. In this paper, 
by taking the fully discrete NLS equation, we firstly constructed its general breather solution via the KP-Toda reduction method. Then by taking a list to the parameters succeeded  in constructing its general rogue wave solution by taking the limit of $p_i \to 0$ successively for $i=1, \cdots, N$. 

We expect to construct rogue wave solutions in other discrete systems such as the discrete complex sine-Gordon equation in the future. 

\section*{Acknowledment}
BF's work is
partially supported by National Science Foundation (NSF) under Grant No. DMS-1715991 and U.S. Department of Defense (DoD), Air Force for Scientific
Research (AFOSR) under grant No. W911NF2010276.


\begin{thebibliography}{99}


\bibitem{akhmediev2009waves}
N. Akhmediev, A. Ankiewicz and M. Taki,
\newblock Waves that appear from nowhere and disappear without a trace.
 {Phys. Lett. A} {\bf 373}, 675--678 (2009).

\bibitem{peregrine1983water}
D.~H.~Peregrine,
\newblock Water waves, nonlinear Schr{\"o}dinger equations and their solutions,
{ J. Aust. Math. Soc. B} {\bf 25} 16 (1983).

\bibitem{akhmediev2009rogue}
N. Akhmediev, A. Ankiewicz, J.~M.~Soto-Crespo.
\newblock Rogue waves and rational solutions of the nonlinear Schr{\"o}dinger equation,
{ Phys. Rev. E}  {\bf 80}, 026601 (2009).

\bibitem{kedziora2012second}
D.~J. Kedziora, A. Ankiewicz and N. Akhmediev,
\newblock Second-order nonlinear Schr{\"o}dinger equation breather solutions in the degenerate and rogue wave limits,
 {Phys. Rev. E}  {\bf 85}, 066601 (2012).


\bibitem{dubard2010multi}
P.~Dubard, P.~Gaillard, C.~Klein and V.~B.~Matveev,
\newblock On multi-rogue wave solutions of the NLS equation and positon solutions of the KdV equation,
{Eur. Phys. J. Special Topics} {\bf 185}, 247 (2010).

\bibitem{dubard2011multi}
P.~Dubard and V.~B.~Matveev,
\newblock Multi-rogue waves solutions to the focusing NLS equation and the KP-I equation,
{ Nat. Hazards Earth Syst. Sci.} {\bf 11}, 667 (2011).

\bibitem{gaillard2011families}
P. Gaillard,
\newblock Families of quasi-rational solutions of the NLS equation and  multi-rogue waves.
{ J. Phys. A: Math. Theor.} {\bf 44}, 435204 (2011).

\bibitem{guo2012nonlinear}
B.~L. Guo, L.~M. Ling and Q.~P.~Liu,
%\newblock Nonlinear Schr{\"o}dinger equation: Generalized darboux transformation and rogue wave solutions.
{ Phys. Rev. E} {\bf 85}, 026607 (2012).

\bibitem{ohta2012general}
Y. Ohta and J.~K. Yang,
%\newblock General high-order rogue waves and their dynamics in the nonlinear Schr{\"o}dinger equation,
 { Proc. R. Soc. London. Sect. A} {\bf 468}, 1716 (2012).

 \bibitem{kharif2009rogue}
\text{C. Kharif, E. Pelinovsky, and A. Slunyaev},
\emph{Rogue Waves in the Ocean}
\newblock  (Springer, Heidelberg, 2009).


\bibitem{onorato2013rogue}
\text{M. Onorato, S. Residori, U. Bortolozzo, A. Montina, and F. T. Arecchi},
Rogue waves and their generating mechanisms in different physical contexts,
\newblock {Phys. Rep.} \textbf{528}, 47 (2013).


\bibitem{chabchoub2011rogue}
\text{A. Chabchoub, N. P. Hoffmann, and N. Akhmediev},
Rogue wave observation in a water wave tank,
\newblock {Phys. Rev. Lett.} \textbf{106}, 204502 (2011).


\bibitem{bludov2009matter}
\text{Y. V. Bludov, V. V. Konotop, and N. Akhmediev},
Matter rogue waves,
\newblock {Phys. Rev. A} \textbf{80}, 033610 (2009).


\bibitem{bailung2011observation}
\text{H. Bailung, S. K. Sharma, and Y. Nakamura},
Observation of Peregrine solitons in a multicomponent plasma with negative ions,
\newblock {Phys. Rev. Lett.} \textbf{07}, 255005 (2011).


\bibitem{solli2007optical}
\text{D.R. Solli, C. Ropers, P. Koonath, and B. Jalali},
Optical rogue waves,
\newblock {Nature} \textbf{450}, 1054 (2007).

\bibitem{kibler2010peregrine}
\text{B. Kibler, J. Fatome, C. Finot, G. Millot, F. Dias, G. Genty, N. Akhmediev, and J. M. Dudley},
The Peregrine soliton in nonlinear fibre optics,
\newblock {Nat. Phys.} {\bf 6}, 790 (2010).

 \bibitem{HedNLS2011} Xu, S.W., He, J.S.,Wang, L.H.: The Darboux transformation of the derivative nonlinear Schrödinger equation.
J. Phys. A {\bf 44}, 305203 (2011)

 \bibitem{GuodNLS2013}  Guo, B.L., Ling, L.M., Liu, Q.P.: High-order solutions and generalized Darboux transformations of derivative
nonlinear Schrödinger equations. Stud. Appl. Math. {\bf 130}, 317--344 (2013)

 \bibitem{ChowdNLS2014}
Chan, H.N., Chow, K.W., Kedziora, D.J., Grimshaw, R.H.J., Ding, E.: Rogue wave modes for a derivative
nonlinear Schrödinger model. Phys. Rev. E {\bf 89}, 032914 (2014)

 \bibitem{HedNLS2017} 
Zhang, Y.S., Guo, L.J., Chabchoub, A., He, J.S.: Higher-order rogue wave dynamics for a derivative nonlinear
Schrödinger equation. Rom. J. Phys. {\bf 62}, 102 (2017)

\bibitem{JiankeJunchaoJNS}
Bo Yang, Junchao Chen, Jianke Yang 
Rogue Waves in the Generalized Derivative Nonlinear Schr\"dinger Equations
Journal of Nonlinear Science, {\bf 30}, 3027--3056 (2020)

\bibitem{BaroniocNLS2012}
Baronio, F., Degasperis, A., Conforti, M.,Wabnitz, S.: Solutions of the vector nonlinear Schrödinger equations:
evidence for deterministic rogue waves. Phys. Rev. Lett. {\bf 109}, 044102 (2012)

\bibitem{BaroniocNLS2014}
Baronio, F., Conforti, M., Degasperis, A., Lombardo, S., Onorato,M.,Wabnitz, S.: Vector rogue waves and
baseband modulation instability in the defocusing regime. Phys. Rev. Lett. {\bf 113}, 034101 (2014)

\bibitem{YYManakov} B. Yang, J. Yang, Universal rogue wave patterns associated with the Yablonskii?Vorob?ev polynomial hierarchy, {\bf 425}, 132958 (2021).


\bibitem{OhtaYangDSI} Rogue waves in the Davey-Stewartson I equation Y Ohta, J Yang
Physical Review E {\bf 86}, 036604 (2012)

\bibitem{OhtaYangDSII} Y Ohta, J Yang,  Dynamics of rogue waves in the Davey-Stewartson II equation, Journal of Physics A {\bf 46}, 105202 (2013)


\bibitem{JiankeBoIMA}
Bo Yang, Jianke Yang 
General rogue waves in the three-wave resonant interaction systems,
IMA Journal of Applied Mathematics, {\bf 86}, 378-425 (2021).

%\bibitem{YYNLSPhysD2021} B. Yang, J. Yang, Rogue wave patterns in the nonlinear Schrödinger equation, Physica D 419 (2021) 132850.

\bibitem{JiankeBoussinesq} B. Yang, J. Yang, General rogue waves in the Boussinesq equation, J. Phys. Soc. Japan {\bf 89}, 024003 (2020)


\bibitem{chen2015rational}
J.~C. Chen, Y. Chen, B.~F. Feng and K. Maruno,
\newblock Rational solutions to two-and one-dimensional multicomponent Yajima-Oikawa systems,
 {Phys. Lett. A} {\bf 379}, 1510 (2015).
%%%%%%%

\bibitem{chen2017YORW}
Junchao Chen, Yong Chen, Bao-Feng Feng,Ken-ichi Maruno, Yasuhiro Ohta, General high-order rogue wave of the (1+1)-dimensional Yajima-Oikawa system, J. Phys. Soc. Jpn, {\bf 87}, 094007
 (2018)

 \bibitem{ankiewicz2010discrete}
A. Ankiewicz, N. Akhmediev and J.~M. Soto-Crespo,
\newblock Discrete rogue waves of the Ablowitz-Ladik and Hirota equations.
 {Phys. Rev. E} {\bf 82}, 026602 (2010).


\bibitem{ohta2014general}
Y. Ohta and J.~K. Yang,
\newblock General rogue waves in the focusing and defocusing Ablowitz-Ladik equations,
 { J. Phys. A: Math. Theor.} {\bf 47} 255201 (2014).

\bibitem{AL1}
 M.J. Ablowitz and J.F. Ladik, A nonlinear dierence scheme and inverse scattering, Stud. in
Appl. Math. {\bf 55}, 213--229 (1976)
 
\bibitem{AL2} M.J. Ablowitz and J.F. Ladik, On the solution of a class of nonlinear partial difference equations, Stud. in Appl. Math. {\bf 57}, 1--12 (1977).

\bibitem{Dis-NLS}  R. Hirota and Y. Ohta, Discrete nonlinear Schr\"odinger equation, talk delivered at
spring meeting of the Physical Society of Japan (1991). Abstract available online:
http://ci.nii.ac.jp/naid/110001908012 (in Japanese).

%\bibitem{APTbook} M.J. Ablowitz, B. Prinari and A.D. Trubach, Discrete and continuous nonlinear Schr\"odinger
%systems, London Mathematical Society Lecture Note Series Vol. 302 (Cambridge University
%Press, Cambridge, 2004).

\bibitem{Dis-NLS-Tsujimoto1} S. Tsujimoto, Y. Ohta and R. Hirota, Difference scheme of nonlinear Schr\"odinger equation,
Proceedings of the annual meeting of Japan Society of Industrial and Applied Mathematics
(1993), pp.203--204 (in Japanese).
\bibitem{Dis-NLS-Tsujimoto2}  S. Tsujimoto, Discretization of integrable systems, in Applied integrable systems, ed. by Y.
Nakamura (Shokabo, Tokyo, 2000), 1--52 (in Japanese).

\bibitem{ILE2019} 
 S. Hirose, J. Inoguchi, K. Kajiwara, N. Matsuura and Y. Ohta,
  Discrete local induction equation,
  J. Integrable Syst. {\bf 3}  xyz003, 1--43 (2019).
  
\end{thebibliography}
\end{document}